# Active Sourced Wavefield Modeling for Layered Half-Space

Mrinal Bhaumik[a]* and Tarun Naskar[b]

*Corresponding Author

[a] PhD Research Scholar

Department of Civil Engineering

Indian Institute of Technology, Madras

Chennai-600036, India

ORCID: 0000-0001-8906-0971

Email: ce19d757@smail.iitm.ac.in

[b] Assistant Professor

BSB-111A, Department of Civil Engineering

Indian Institute of Technology, Madras

Chennai-600036, India

ORCID: 0000-0002-2752-2840

Email: tarunnaskar@civil.iitm.ac.in

**ACKNOWLEDGMENTS**

The authors acknowledge the financial support provided by the Science and Engineering Research Board (SERB), Department of Science & Technology, Government of India. The project's reference no. CRG/ 2022/004112.

**ABSTRACT**

Traditional free vibration-based forward models generate theoretical dispersion curves under the assumption of planar waves, neglecting the influence of the actual source-receiver configuration. While 2D/3D numerical wavefield modeling approaches mimic real-field scenarios with source-receiver information, they suffer from computational inefficiency. This study introduces a semi-analytical wavefield modeling approach incorporating source-receiver data acquisition layouts. The method considers a cylindrically spreading wavefield described by the Hankel function instead of the planer wave assumption. The approach considers both propagating waves characterized by real wavenumbers and decaying waves with complex wavenumbers, allowing for calculating surface displacements in both the far and near fields. The proposed model captures the complete wavefield, including source-offset effects and leaky waves, while maintaining computational efficiency comparable to any free vibration-based approaches. The method entails solving the eigenvalue problem constructed through the higher-order thin-layer method. Subsequently, it calculates the frequency domain vertical and radial surface responses at any desired location in space generated by a vertically positioned active source. The overall performance of the proposed method is investigated on diverse profiles, including regularly dispersive media, low-velocity layer models, and thin plate structures. The vertical and radial component dispersion images are validated against the numerical approach. The proposed method is at least two orders of magnitude faster than the numerical method. Notably, it captures the smooth transition of modal energy from the fundamental mode to higher modes, occurring due to modal osculation at low frequencies. The present approach offers a valuable tool to enhance the efficiency of active surface wave methods.

## 1 INTRODUCTION

Wavefield modeling has wide-ranging applications across various engineering fields, including non-invasive geotechnical site investigation, near-surface geophysics, pavement quality monitoring, and the non-destructive evaluation of structural elements. The common 1D forward modeling approaches include the transfer matrix method [1], stiffness matrix method [2,3], Schwab and Knopoff method [4,5], Kennet's reflection-transmission method [6], and thin-layer method (TLM) [7,8]. These methods primarily rely on the assumption of planar Rayleigh wave propagation and do not incorporate actual source-receiver layouts. Consequently, they yield only theoretically possible modes without providing information about their relative spectral contributions in surface wave dispersion. The ambient noise-based passive seismic methods are likely to hold this assumption reasonable as the source location is unknown and anticipated to be far from the receiver location. However, for active source-based applications, the dispersion images are sensitive to data acquisition parameters such as spread length, source offset, and source energy [9–11]. Dispersion curves extracted with varying source offsets yield contrasting results, introducing uncertainty into surface wave inversion analyses [12]. 2D/3D discretization-based numerical forward modeling approaches, including finite element (FE), spectral element (SE), finite difference (FD), and staggered grid finite difference (SGFD) methods, solve the governing partial differential equations and provide seismograms or dispersion images that closely resemble field data [13,14]. Numerical modeling considers actual field scenarios, such as source-receiver layout, near-field effects, and mode jump. However, numerical approaches are computationally intensive and take several minutes to a couple of hours to solve a single forward model. The fundamental reason behind increased computational time is due to the Courant stability condition: the length of the time step must be smaller than the travel time of maximum speed across any two grids in space[15]. Additional complexities with numerical methods involve accurately modeling free surface conditions and designing perfect absorption boundary layers at discontinuous edges. Nevertheless, discretization-based numerical modeling is advantageous only in simulating complex profiles, including lateral heterogeneity. In recent wavefield modeling advancements for horizontally stratified media, Bhaumik and Naskar [16] introduced a higher-order thin layer method (HTLM) based approach, producing a full velocity spectrum considering the field acquisition layout. This method demonstrates computational efficiency, being at least two orders faster than numerical approaches, and is insensitive to high Poisson's ratios. However, it is crucial to note that this approach is grounded on the planar wave assumption, which is valid only at the far field from the source.

The present study proposes an active sourced semi-analytical wavefield modeling approach considering cylindrical wavefront in the form of Hankel function instead of planer wave assumption. Additionally, the method accounts for both propagating and evanescent modes, enabling the calculation of surface displacements in far and near fields. It generates vertical and radial component Rayleigh wave dispersion images similar to that produced by 2D numerical approaches while maintaining the computational speed comparable to the 1D free vibration-based methods. The present method is capable of modeling the complete wavefield, encompassing body waves, surface waves, and leaky waves. In contrast to the 2D numerical approaches, an analytical solution is used along the horizontal direction, and finite element discretization in the vertical direction. Similar to Bhaumik and Naskar [16], the present method utilizes higher-order thin-layer techniques to compute element stiffness matrices. The use of a higher-order interpolation function reduces the number of thin layers, thereby lowering computational costs. Unlike numerical approaches, the proposed method calculates solutions only at defined receiver positions, significantly reducing the computational burden of calculating the displacement at all the element nodes. The frequency domain complex vertical and radial displacement responses are recorded at desired locations on the surface. Finally, spectral values for each frequency are explored over a wide range of trial phase velocities to obtain the dispersion spectrum. To demonstrate the accuracy of the proposed method, four distinct geological profiles are selected, representing regularly dispersive media, an embedded low-velocity layer model, and a crustal-scale model. Additionally, the procedure is extended to model Lamb waves propagating through plate-like structures. The phase velocity spectrum from the proposed approach is compared with a 2D numerical model to validate the derived formulation. The results across different soil profiles showcase the effectiveness of the proposed framework in modeling intricate phenomena such as mode jump and mode osculation. It adeptly captures the smooth transition of modal energy from the fundamental mode to higher modes at low frequencies due to the presence of bedrock, addressing the challenge of mode misidentification and overestimation of shear wave velocity during inversion. Moreover, it can capture the downward trend of phase velocities at low frequencies due to near-field effects when receivers are positioned close to the source. Given the computational time of our proposed method is comparable to any free vibrations-based Rayleigh wave mode computing approaches, it can potentially enhance inversion algorithms that consider the entire spectrum.

## 2 METHODOLOGY

The derivation of dynamic responses in laterally homogeneous stratified media commences with obtaining the eigenvalues and eigenvectors of the surface wave modes. Existing methods for calculating wave modes include secular function-based methods such as transfer matrix [1], dynamic stiffness matrix [2], Schwab and Knopoff method [4], R-T coefficient method [6], and discretization methods such as TLM [2,17] and HTLM[16]. The TLM is widely utilized for wave propagation modeling in layered structures [3,16,18,19]. It formulates the problem as a conventional quadratic eigenvalue problem, unlike the transcendental form used by root search methods. Additionally, TLM enables both propagating and decaying modes with complex eigenvalues and excels in modeling damping effects within the viscoelastic medium. The HTLM encompasses all the benefits of TLM while offering an additional advantage of enhanced computational efficiency attributed to the utilization of higher-order shape functions. Other 1D discretization methods, such as finite-difference [20], pseudo-spectral [21], and spectral element method [22], can also be employed for modal analysis. In the present study, we followed the HTLM method described in Bhaumik and Naskar [16].

### 2.1 Formulation of the Eigenproblem

Consider a laterally homogeneous layered half-space system shown in Fig.1a, with boundaries at $z = 0$ and $z = l$ referred to as free surface and elastic half-space, respectively. In the absence of body force and damping, the elastodynamics equation can be written as,

$$\nabla^T \boldsymbol{\sigma} + \rho_s \omega^2 \mathbf{u} = \mathbf{0} \tag{1}$$

where, $\boldsymbol{\nabla} = [\partial/\partial x \ 0; \ 0 \ \partial/\partial z; \ \partial/\partial z \ \partial/\partial x]$ is differential operator, $\boldsymbol{\sigma} = \{\sigma_{xx} \ \sigma_{zz} \ \sigma_{xz}\}^T$ is stress vector, $\rho_s$ is the density of the layer, $\omega \in \mathbb{R}$ is frequency, and $\mathbf{u} = \{u_x \ u_z\}^T$ is displacement vector. Using the analytical solution in the form $u_x(x,z) = U(z)e^{i(\omega t - kx)}$ and $u_z(x,z) = W(z)e^{i(\omega t - kx)}$, and applying finite element discretization in the vertical direction, the following eigenvalue problem can be obtained [2]:

$$[k^2 \mathbf{A} + ik\mathbf{B} + (\mathbf{C} - \omega^2 \mathbf{M})] \begin{Bmatrix} \mathbf{U} \\ \mathbf{W} \end{Bmatrix} = 0 \tag{2}$$

where $\mathbf{A}, \mathbf{B}, \mathbf{C},$ and $\mathbf{M}$ are global stiffness matrices obtained by assembling all the layer stiffness matrices, $k$ is the spatial wavenumber, $\mathbf{U}$ and $\mathbf{W}$ are vectors with nodal displacements along the horizontal and vertical directions. The derivation of layer matrices is extensively discussed by Kausel and Roësset [2]. Bhaumik and Naskar [16] implemented higher-order shape functions to derive the element stiffness matrices, with a brief overview provided in Appendix A for completeness.

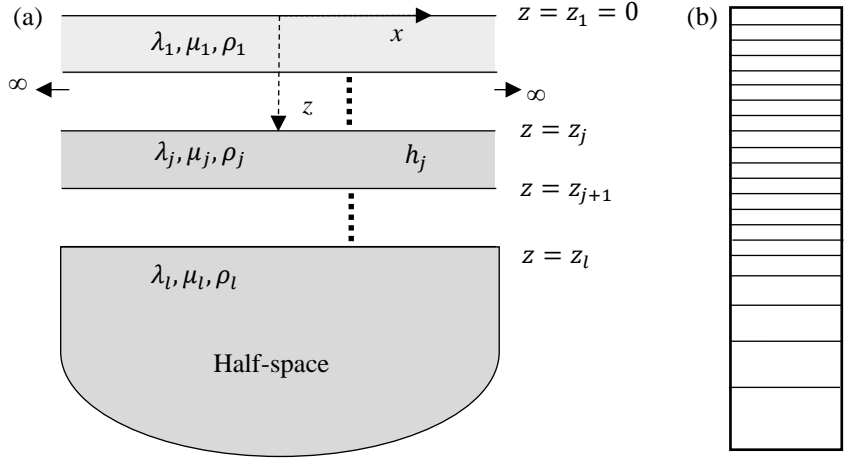

**Fig. 1.** (a) Layered half-space model (b) discretization of finite layers and infinite half-space.

The elastic half-space is modeled using perfectly matched discrete layers (PMDL). The concept of impedance-preserving properties of the midpoint integrated linear element is implemented in developing PMDL elements [23]. Vaziri Astaneh and Guddati [24] investigated the performance of PMDL elements in the forward modeling of the Rayleigh wave. The adequately chosen element length along with mid-point integration, can model the exact elastic half-space with a few elements. A simplified procedure to calculate the element thickness [24]:

$$h_L = h_1 \alpha^{L-1} \quad (L = 2 \ldots \bar{L}); , \tag{3}$$

where $\alpha$ is the geometric ratio and $\bar{L}$ is the number of half-space layers elements. A schematic representation of the discretized profile, illustrating the varying thickness of the elements, is provided in Fig. 1b. For more efficient modeling of complex-valued evanescent modes, PMDL layers with complex lengths can be employed. The half-space formulation is ignored when modeling Lamb waves through plate-like structures.

Once the element stiffness matrices for the finite layer and half-space have been calculated, the quadratic eigenvalue problem in Equation 2 can be transformed into a generalized eigenvalue problem [2]:

$$\left(k^2 \begin{bmatrix} \mathbf{A}_{xx} & 0 \\ \mathbf{B}_{xz}^T & \mathbf{A}_{zz} \end{bmatrix} + \begin{bmatrix} \mathbf{G}_{xx} & \mathbf{B}_{xz} \\ 0 & \mathbf{G}_{zz} \end{bmatrix}\right) \begin{Bmatrix} \mathbf{U} \\ ik\mathbf{W} \end{Bmatrix} = \mathbf{0} \tag{4}$$

where the $\mathbf{A}_{xx}$ and $\mathbf{A}_{zz}$ are submatrices of global stiffness matrix $\mathbf{A}$ arranged according to the degree of freedom, as discussed in Appendix A. Similarly, $\mathbf{B}_{xz}$, $\mathbf{G}_{xx}$ and $\mathbf{G}_{zz}$ are submatrices of the global stiffness matrix given in Equation 2. The eigenvectors obtained from the solution of the generalized eigenvalue problem are normalized to satisfy the modal orthogonality condition [16,19]. The normalized eigenvector, $\boldsymbol{\phi} = \mathbf{R}\boldsymbol{\chi}^{-1/2}$, where $\boldsymbol{\chi} = \mathbf{L}^T\mathbf{A}\mathbf{R}\boldsymbol{\Lambda}^{-1/2}$, $\mathbf{R}$ is the right eigenvector, given by $\mathbf{R} = [\mathbf{U} \quad ik\mathbf{W}]^T$, left eigenvector $\mathbf{L} = [k\mathbf{U} \quad i\mathbf{W}]$, and $\boldsymbol{\Lambda} = diag\{k_1^2, k_2^2, \ldots\}$ is diagonal vector contains the eigenvalues. The normalized eigenvectors are then used in the computation of the dynamic surface responses.

### 2.2 Calculation of Surface Displacements

Surface waves generated from a point source on the surface spread cylindrically (Fig. 2). Assuming a vertically applied disk-like load (Fig.3) of intensity $p$ at the $s$th interface of a thin layered model, the displacement of the $l$th Rayleigh wave mode at a distance $r$ and at the $q$th interface is given by [8]:

$$d_v^{qm}(r, \omega) = pR \frac{\pi}{i2k_m} \phi_z^{sm} \phi_z^{qm} J_1^{(1)}(k_m R) H_0^{(2)}(k_m r) \quad R \leq r \tag{5}$$

where $R$ is the radius of disk load; $k_m$ is the wavenumber of $m$th mode at frequency $\omega$; $\phi_z^{sm}$ and $\phi_z^{qm}$ are the $m$th mode normalized vertical component eigenvector at $s$th interface (source location) and $q$th interface (receiver location), respectively; $J_1^{(1)}(k_m R)$ is the Bessel function of 1st kind of 1st order and $H_0^{(2)}(k_m r)$ is the Hankel function of 2nd kind of zero order.

In surface wave testing, the wavefield observed at the surface is a result of the superposition of multiple modes. When both the load and receiver are positioned at the surface ($s = 1, q = 1$), the equation for the vertical displacement of surface waves at a given frequency, $\omega$, can be expressed as:

$$d_v^{surf}(r,\omega) = \sum_{m=1}^{M} d_v^{1m}(r,\omega) = \sum_{m=1}^{M} pR \frac{\pi}{i2k_m} \phi_z^{1m} \phi_z^{1m} J_1^{(1)}(k_m R) H_0^{(2)}(k_m r) \qquad (6)$$

where, $M$ is the number of modes.

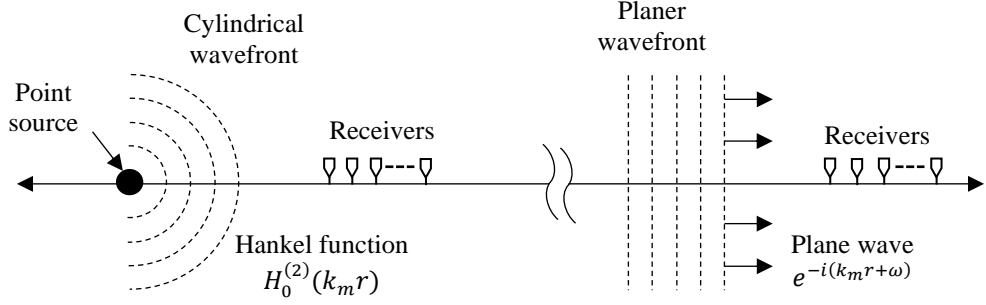

**Fig. 2**. Illustration of cylindrical and planer waveform.

Solving the eigenvalue problem in Equation 4 yields two sets of eigenvalues. Half of these eigenvalues, with positive real and negative imaginary wavenumbers, represent waves that are propagating and decaying away from the source [17]. In contrast, the other half of the eigenvalues represent waves that are traveling towards the source. To accurately calculate the displacement response near the source, both the propagating waves with real wavenumbers and the decaying waves with complex wavenumbers (having a negative imaginary part) need to be considered [19]. It is important to note that the simple mode summation approach, which considers only the real possible roots, fails to accurately model the transient near-field effects. It does not account for the waves that decay as they travel away from the source, which play a significant role in the near-field region. By including both propagating and decaying waves in the calculation, the proposed approach provides a more accurate representation of the displacement field, especially in the vicinity of the source.

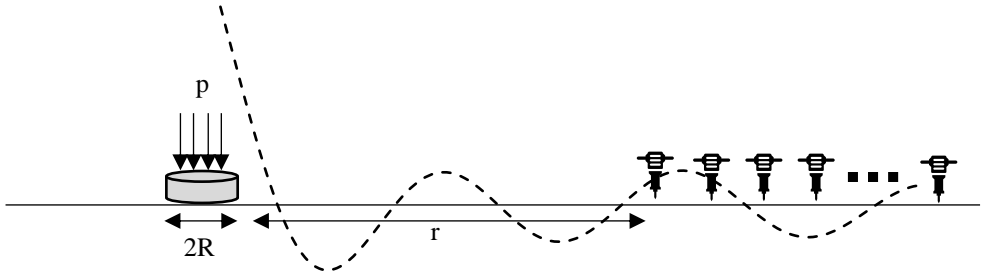

**Fig. 3.** Displacement response due to circular plate loading.

Equation 6 shows that the modes propagate in the form of Hankel functions rather than as plane wavefronts. The Hankel function can be expressed as, $H_0^{(2)}(k_m r) = J_0^{(1)}(k_m r) - i J_0^{(2)}(k_m r)$, where $J_0^{(1)}(k_m r)$ and $J_0^{(2)}(k_m r)$ are Bessel function of 1st and 2nd kind, respectively. In the far field, where $k_m r \gg 1$, the zero-order Bessel function of 1st and 2nd kind can be approximated as:

$$J_0^{(1)}(k_m r) \approx \sqrt{\frac{2}{\pi k_m r}} \cos\left(k_m r - \frac{\pi}{4}\right)$$

$$J_0^{(2)}(k_m r) \approx \sqrt{\frac{2}{\pi k_m r}} \sin\left(k_m r - \frac{\pi}{4}\right) \qquad (7)$$

So, the Hankel function $H_0^{(2)}$ in Equation 6 becomes:

$$H_0^{(2)}(k_m r) = \sqrt{\frac{2}{\pi k_m r}}\left[\cos\left(k_m r - \frac{\pi}{4}\right) - i\sin\left(k_m r - \frac{\pi}{4}\right)\right] = \sqrt{\frac{2}{\pi k_m r}} e^{-i(k_m r - \frac{\pi}{4})} \tag{8}$$

Now, at the far field, Equation 6 transforms to

$$d_{v,far}^{surf}(r,\omega) = \sum_{m=1}^{M} pR \frac{\pi}{i2k_m} \phi_z^{1m}\phi_z^{1m} J_1^{(1)}(k_m R) \sqrt{\frac{2}{\pi k_m r}} e^{-i(k_m r - \frac{\pi}{4})} \tag{9}$$

It is clear that at a far field, modes propagate approximately in the `form of a cylindrical wavefront, and the geometrical decay is proportional to $1/\sqrt{r}$.

Similarly, radial component surface displacement at a distance $r$ due to the same vertical disk loading of intensity $p$ at the surface [8]:

$$d_h^{surf}(r,\omega) = \sum_{m=1}^{M} pR \frac{\pi}{i2k_m} \phi_z^{1l}\phi_x^{1l} J_1^{(1)}(k_m R) H_1^{(2)}(k_m r) \quad R \leq r \tag{10}$$

where, $\phi_x^{1m}$ is the normalized horizontal eigenvector component of $m$th mode at the surface obtained from **U**, and $H_1^{(2)}(k_m r)$ is the Hankel function of second kind of first order. The frequency domain complex valued surface displacement $d(r,\omega)$ contain amplitude and phase information. Therefore, sinusoids with corresponding amplitude and phase for the desired frequency range can be added to obtain the time domain responses at each receiver location.

### 2.3 Computation of Phase Velocity Spectrum

In surface wave methods, the responses are often measured at the desired locations on the surface, as depicted in Fig. 3. To better understand the wave propagation characteristics, it is convenient to represent the kinematics of the propagating modes in an appropriate domain, such as the frequency-wavenumber domain or the frequency-phase velocity domain. The method is often referred to as the wavefield transformation [25]. Using the phase velocity scanning procedure, the spectral energy (**E**) of any trial phase velocity ($c$), at frequency $\omega$:

$$\mathbf{E}(\omega, c) = \sum_{n=1}^{X} \mathbf{d}(r_n, \omega) e^{-i\omega r_n/c} \tag{11}$$

where, $X$ is the number of receivers. Performing this operation for each frequency yields a dispersion image. Note that normalizing $\mathbf{d}(r_n, \omega)$ with the absolute amplitude, the method is similar to the phase shift transform [26]. The amplitude in the dispersion spectrum can be normalized corresponding to the maximum amplitude at each frequency:

$$\bar{\mathbf{E}}(\omega, c) = \frac{|\mathbf{E}(\omega, c)|}{\max(|\mathbf{E}(\omega, c)|)} \tag{12}$$

This process aids in identifying the phase velocity of different modes present in the surface wavefield and enables a better representation of the dispersion image.

### 3  PERFORMANCE OF THE PROPOSED METHOD

To validate and demonstrate the accuracy of the proposed approach, the seismograms and phase velocity dispersion images for four different geological models have been compared with results obtained using 2D numerical simulation. The 2-4 dispersive SGFD (DSGFD) method (second order in time and fourth order in space) is employed as the numerical seismic wavefield modeling tool. The SGFD method is widely used in seismology due to its known accuracy and stability. The DSGFD method takes advantage of the dispersive nature of Rayleigh waves and employs non-uniform discretization along the depth, resulting in reduced computational demands compared to the uniform grid SGFD method [27]. The method numerically solves the first-order velocity-stress formulation of the wave equation through a grid-staggering approach, where velocity and stress components are shifted by a half-grid [13,27]. The spatial and temporal resolution is determined from the numerical stability criteria. The spatial discretization is done using 20 grid points per minimum wavelength, and the perfectly matched layer (PML) boundary condition is applied along the three discontinuous edges. The vertical and radial component seismograms and dispersion images generated by the proposed method were compared to those obtained using the DSGFD method while maintaining an identical source-receiver layout.

**Table 1.** Layer parameters of adopted soil models

| Sl. No | Profile | Thickness (m) | Shear wave velocity (m/s) | Compression wave velocity (m/s) | Poisson's ratio | Density (kg/m$^3$) |
|---|---|---|---|---|---|---|
| 1 | Profile: I | 10 | 200 | 800 | 0.467 | 2000 |
|  |  | Half-space | 400 | 1200 | 0.437 | 2000 |
| 2 | Profile: II | 2 | 194 | 650 | 0.45 | 1820 |
|  |  | 2.3 | 270 | 750 | 0.425 | 1860 |
|  |  | 2.5 | 367 | 1400 | 0.46 | 1910 |
|  |  | 2.8 | 485 | 1800 | 0.46 | 1960 |
|  |  | 3.2 | 603 | 2150 | 0.457 | 2020 |
|  |  | Half-space | 740 | 2800 | 0.46 | 2090 |
| 3 | Profile: III | 10 | 250 | 496.3 | 0.33 | 2000 |
|  |  | 10 | 150 | 297.8 | 0.33 | 2000 |
|  |  | Half-space | 300 | 595.6 | 0.33 | 2000 |
| 4 | Profile: IV | 1000 | 3800 | 7544 | 0.33 | 2700 |
|  |  | 1000 | 3500 | 6948 | 0.33 | 2500 |
|  |  | 1000 | 4400 | 8735 | 0.33 | 3100 |
|  |  | 1000 | 4100 | 8139 | 0.33 | 2900 |
|  |  | Half-space | 4700 | 9330 | 0.33 | 3300 |

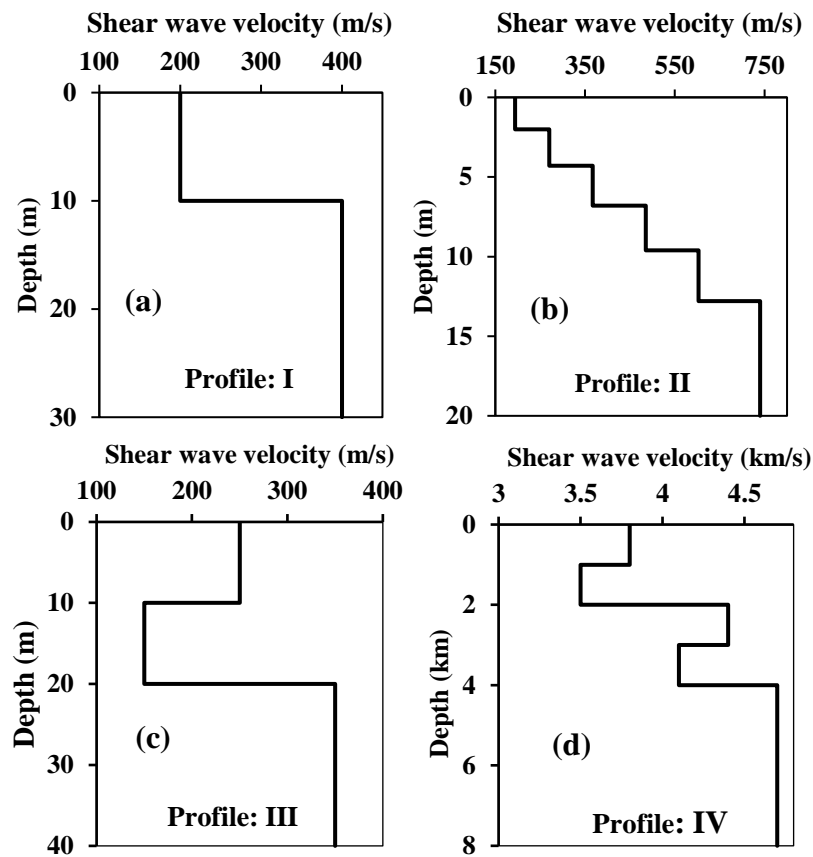

**Fig. 4**. Four representative synthetic soil profiles

Four representative synthetic soil profiles are shown in Fig.4. The detailed parameters of the models are provided in Table 1. Each model reflects different field scenarios commonly encountered in near-surface investigations, including both normally dispersive profiles and velocity reversal. Profile I consists of a single layer over an elastic half-space, selected from Xu et al [28]; Profile II represents a normally dispersive profile with significant velocity contrast at deeper depths from Xia et al [29]; Profile III is an irregularly dispersive model used to study higher modes [30]; and Profile IV is a crustal level model, found in Naskar and Kumar[31]. In addition to the soil models, the analysis also includes a concrete plate structure to model Lamb waves.

## 3.1 Comparison of Dispersion Spectrum

### 3.1.1 Profile I: Two-layer model

A simple two-layered regularly dispersive model is considered first (Profile I). The upper layer, with a thickness of 10 m and a Poisson's ratio of 0.47, represents a soft soil stratum. The top layer is discretized into 10 thin sub-layers, and quartic interpolation functions are employed to compute the stiffness matrices. The underlying half-space is modeled using 10 mid-point integrated linear PMDL elements. With the proposed approach, the vertical and radial displacements in the frequency domain are calculated at 48 locations on the surface, spanning a frequency range of up to 50 Hz, with a frequency resolution of 0.5 Hz. For the DSGFD simulation, a wavefield of dimensions 110 m × 50 m is discretized into 0.2 m grid elements. The seismic source is modeled using a 20 Hz Ricker wavelet.

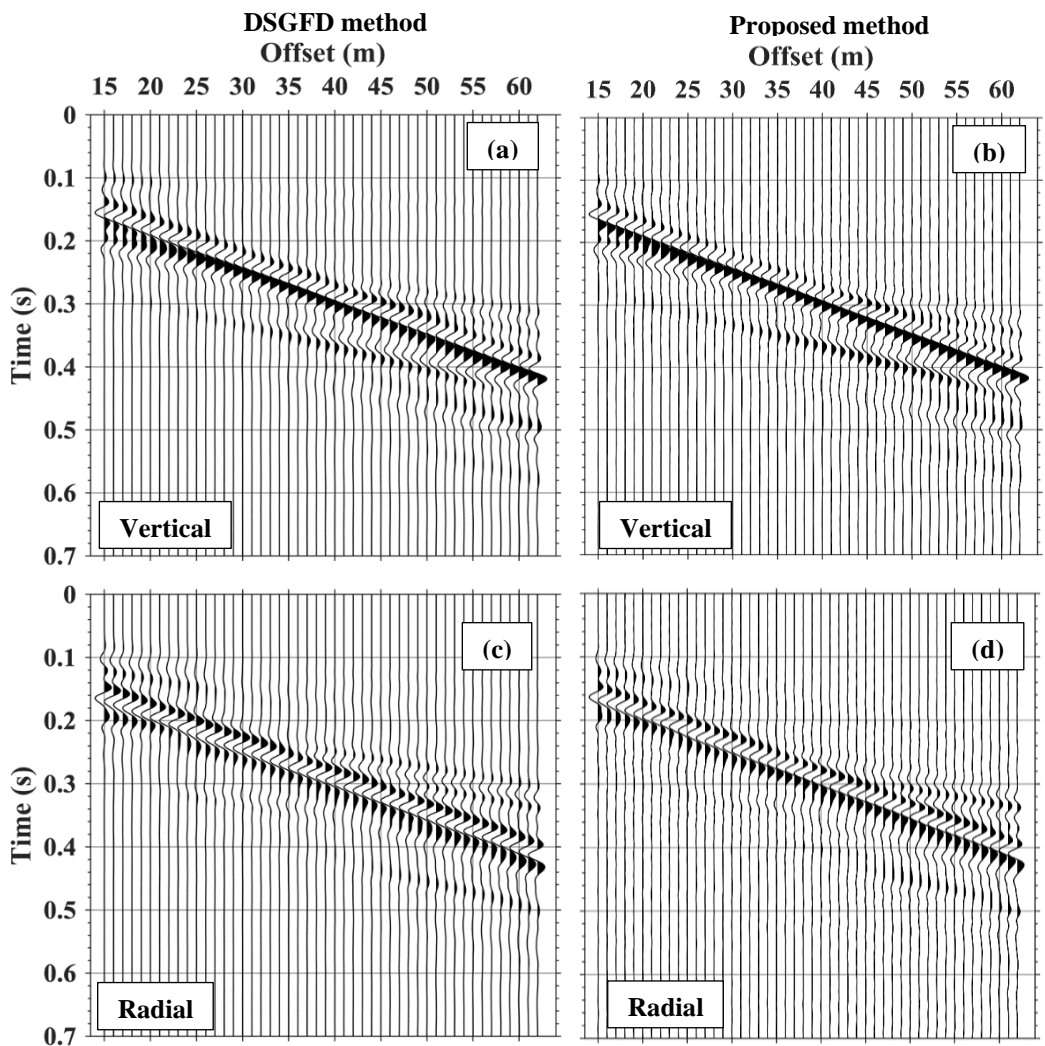

**Fig. 5.** Comparison of seismograms obtained using proposed method and DSGFD method for Profile I: (a) vertical component from DSGFD method, (b) vertical component from proposed method, (c) radial component from DSGFD method, and (d) radial component from the proposed method.

The vertical and radial component surface responses computed from the proposed method are compared with the reference seismograms of the DSGFD method (Fig. 5). The wave trends calculated using the proposed approach closely match those obtained from the numerical solution, accurately capturing the presence of refracted and reflected surface and body waves. To compare the responses in the frequency domain, the vertical and radial component dispersion images are plotted in Fig. 6. Both methods yield identical dispersion images, confirming the consistency and accuracy of the proposed approach. The white dotted line denotes the theoretical modes. As anticipated, the fundamental mode energy dominates the entire frequency range. The 1st and 2nd higher modes exhibit more potency in the vertical component dispersion image compared to the radial one. While the theoretical modes are invariant irrespective of the direction of the component, the possible difference between the vertical and radial components of the Rayleigh wave offers an additional constraint for inversion analysis. The anomalies below 8 Hz observed in the horizontal component dispersion image obtained from the numerical approach are attributed to numerical discretization errors. In contrast, the present method operates in the frequency domain, where the use of higher-order discretization in the vertical direction helps to avoid such errors.

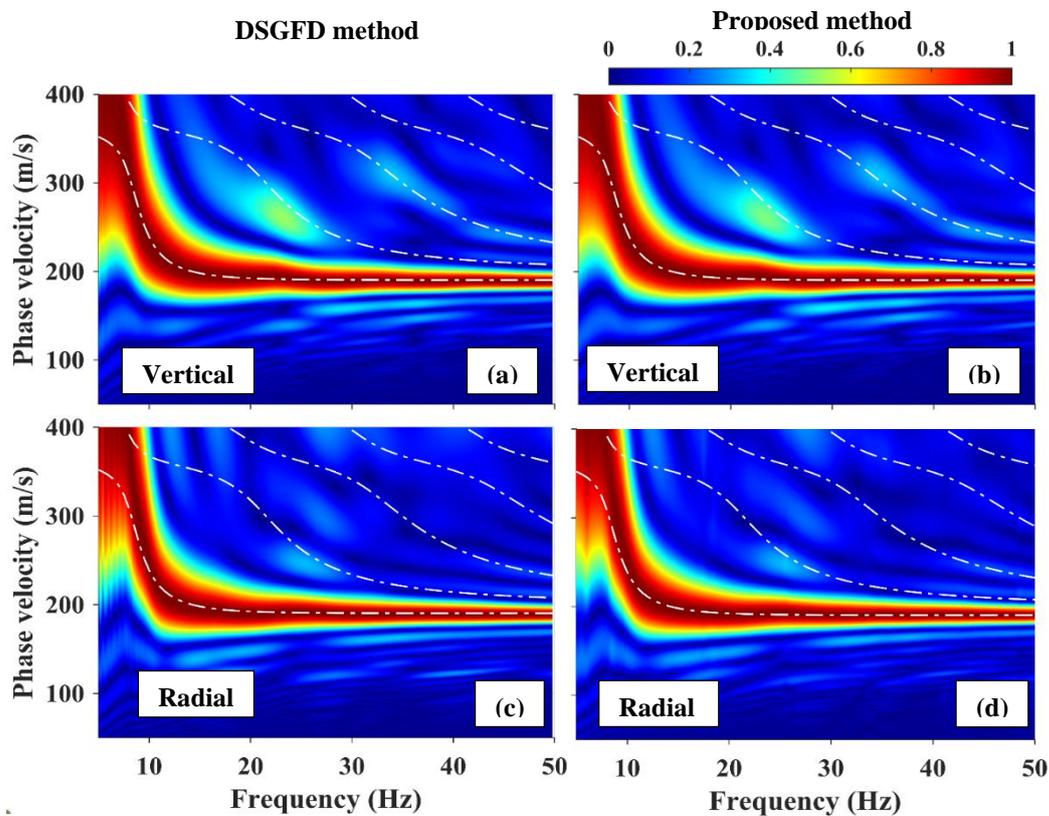

**Fig. 6**. Comparison of dispersion images obtained using proposed method and DSGFD method for Profile I: (a) vertical component from DSGFD method, (b) vertical component from proposed method, (c) radial component from DSGFD method, and (d) radial component from the proposed method.

*3.1.2  Profile II: Six-layer normally dispersive model*

Profile II represents a regularly dispersive site with a significant impedance contrast between the surface layer and half-space (Table 1). This scenario is frequently encountered in real-world geological profiles with underlying bedrock. The sharp impedance contrast leads to the excitation of higher modes in the low-frequency region, a well-known behavior for causing mode misidentification and the overestimation of shear wave velocity. Therefore, Profile II was deliberately chosen to evaluate the capability of the proposed method to capture and represent this phenomenon accurately.

The layer stiffness matrices are calculated using quartic interpolation functions, and 10 mid-point integrated linear PMDL layers are employed to model the elastic half-space. A 140 m × 40 m model is utilized for numerical modeling with a grid spacing of 0.1 m. The vertical and radial component surface responses are recorded at 48 locations spaced 1 meter apart. Fig. 7 presents seismograms obtained through the proposed approach and the DSGFD method. The overall waveform for both components closely agree with the numerical results. The dominance of Rayleigh waves is observed in the vertical component, while the radial component exhibits significant energy from body waves. Dispersion images of the seismograms are compared in Fig. 8. Both the vertical and radial component dispersion images obtained from the proposed method closely resemble the results of the DSGFD method. When plotting the theoretical dispersion curves (white dot lines), it becomes apparent that the fundamental and the first higher modes are closely located around 15 Hz. This occurs due to the high-velocity contrast in deep layers, leading to mode osulation, where modal energy shifts from the fundamental mode to higher modes at low frequencies. In the dispersion image, the dominant modal energy shifts to the higher mode at frequencies lower than the osulation frequency. However, due to the limited number of receivers in field acquisition, mode jumping at low frequencies in the field dispersion image is often poorly detected. Consequently, the higher mode at low frequency is frequently misidentified as the fundamental mode, resulting in an overestimation of shear wave velocity during inverse analysis. However, instead of relying solely on theoretical dispersion curves, utilizing the full dispersion image generated by the proposed method can help avoid this problem. Additionally, the energy distribution in higher modes in the radial component dispersion image differs from that in the vertical component dispersion image. The radial component dispersion image displays a higher spectral energy concentration for the first higher mode between 20-40 Hz. Therefore, incorporating the radial component dispersion image adds valuable constraints to the inversion analysis.

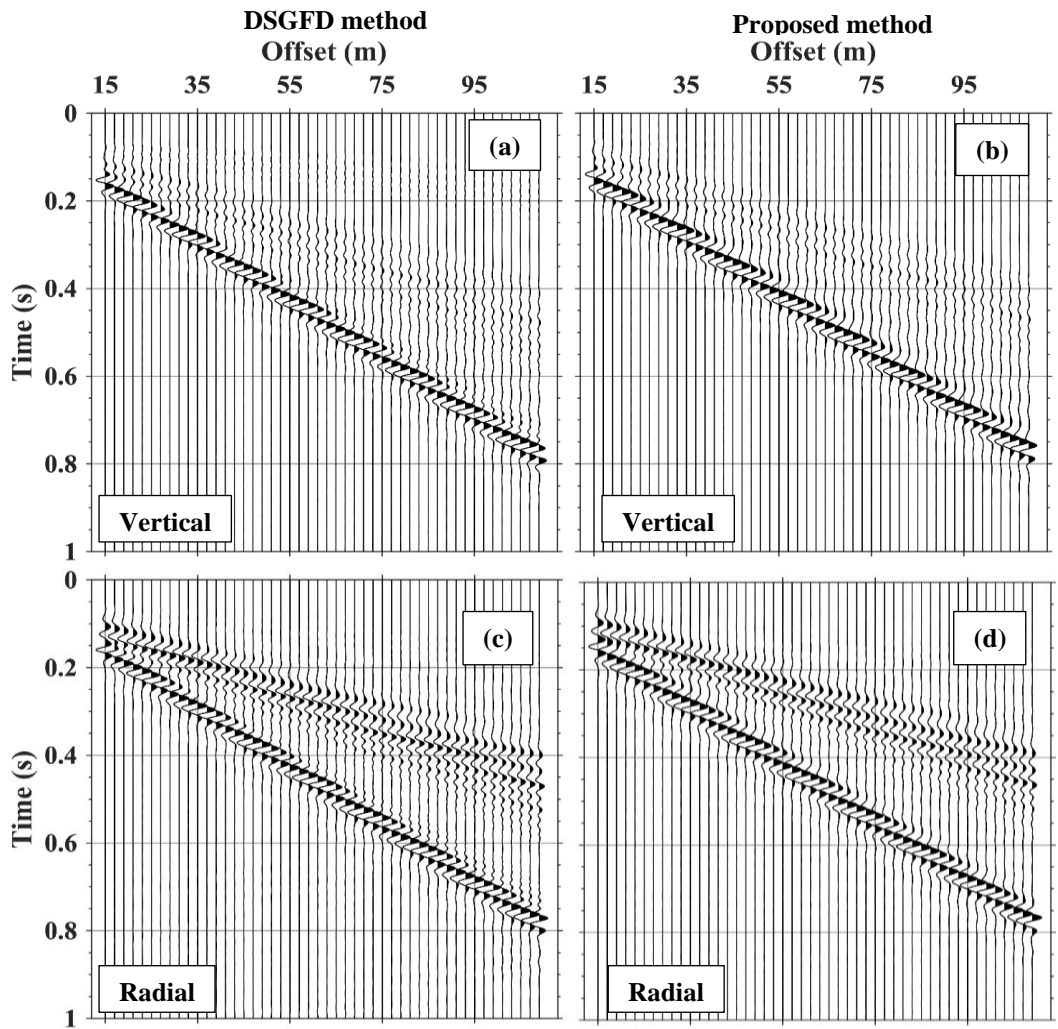

**Fig. 7.** Comparison of seismograms obtained using proposed method and DSGFD method for Profile II: (a) vertical component from DSGFD method, (b) vertical component from proposed method, (c) radial component from DSGFD method, and (d) radial component from the proposed method

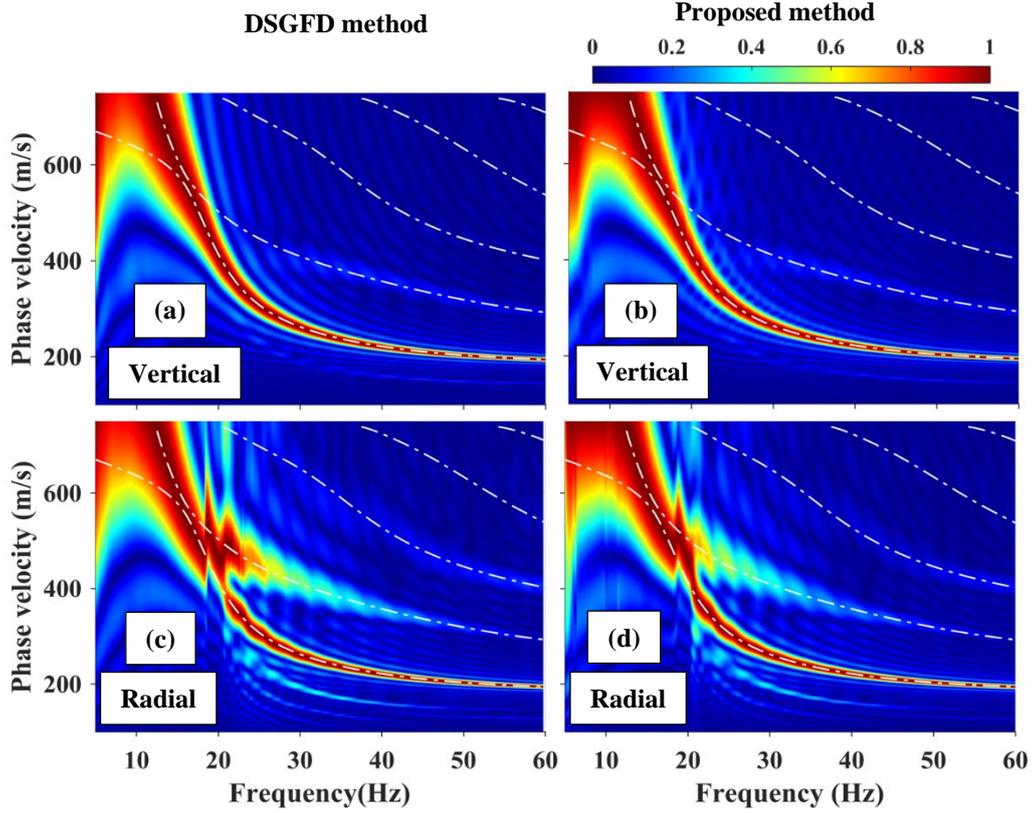

**Fig. 8.** Comparison of dispersion images obtained using proposed method and DSGFD method for Profile II: (a) vertical component from DSGFD method, (b) vertical component from proposed method, (c) radial component from DSGFD method, and (d) radial component from the proposed method.

*3.1.3    Profile III: Irregularly dispersive profile*

Profile III represents an irregularly dispersive media where a stiff layer overlays a soft layer (Table 1). Such complex soil layers are frequently encountered in the field, particularly where the top surface has become relatively stiffer due to soil desiccation. Unlike Profile I, where only the fundamental mode is dominant across the entire frequency range, irregularly dispersive media often exhibit the significant influence of higher modes within specific frequency ranges. Consequently, an irregularly dispersive profile has been selected to demonstrate the effectiveness of the proposed method in simulating higher modes and the occurrence of modal energy jumps. Each layer has been discretized into 2-meter-thick sub-layers, and the stiffness matrices have been calculated using quartic interpolation functions. The elastic half-space is modeled using ten mid-point integrated linear PMDL elements. In the DSGFD method, a 135 m × 40 m wavefield is spatially discretized using 0.2 m elements. The vertical and radial component responses were calculated at 48 locations on the surface, spaced 2 meters apart. A larger spread length is selected to achieve a dispersion image with well-separated higher modes. Fig. 9 illustrates a close match between the vertical and radial component seismograms obtained by the proposed approach and the DSGFD method. A clear presence of higher modes is observed in the seismogram from a distance of 65 m onwards from the source. The dispersion images for both vertical and radial components are compared with the results obtained through the DSGFD method in Fig. 10. The fundamental mode dominates up to 10 Hz in both components, and clear occurrences of modal energy jumps are observed at different cut-off frequencies. The dispersion images of both the vertical and radial components obtained through the proposed approach closely match those obtained from the DSGFD method. This profile showcases the proposed method's capability to model higher modes effectively and accurately capture modal energy transitions.

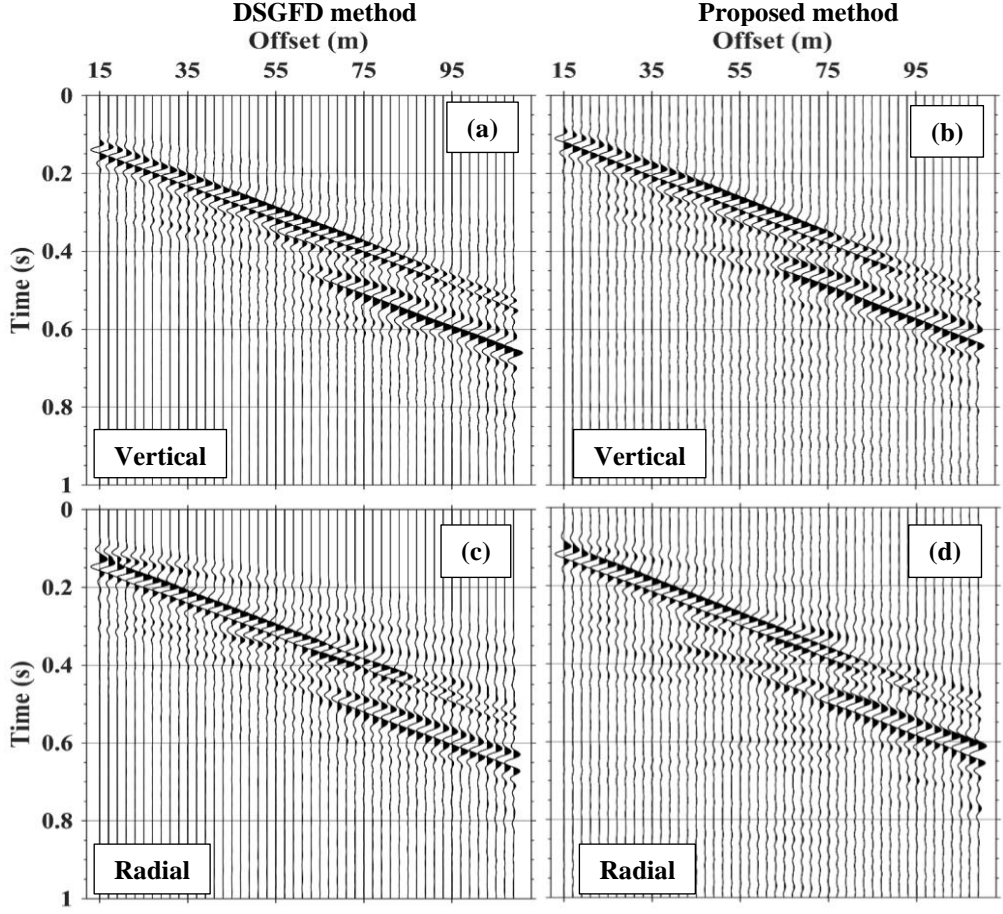

**Fig. 9.** Comparison of seismograms obtained using proposed method and DSGFD method for Profile III: (a) vertical component from DSGFD method, (b) vertical component from proposed method, (c) radial component from DSGFD method, and (d) radial component from the proposed method.

*3.1.4   Profile IV: Crustal model*

This example considers a crustal level profile consisting of four layers of each 1 km thickness (Fig.4d). Profile IV has two low-velocity layers sandwiched between the stiff layers. The crustal-scale profile is useful to demonstrate the effectiveness of the present approach in the low-frequency range. The detailed parameters are described in Table 1. In the proposed approach, each of the 1 km layers is discretized into 0.5 km sub-layers, and a quartic interpolation function is used to construct the element stiffness matrices. 15 PMDL layers are employed to model very low frequencies. Vertical and radial component responses are recorded at 60 locations on the surface, spaced at 0.5 km intervals. In the DSGFD method, a 33 km × 15 km wavefield is discretized with 20 m elements. Fig.11 compares the seismograms recorded using the DSGFD method and the proposed method. The overall wave trends for both components obtained by the proposed method are similar to those obtained with the DSGFD method. To further analyze the results in the frequency domain, the dispersion images are presented in Fig. 12. Both the vertical and radial component dispersion images obtained by the proposed method match well with the results of the 2D numerical method. A hump in the dispersion image near 1 Hz frequency is due to the low-velocity layer. This comprehensive analysis validates the overall accuracy and reliability of the proposed method.

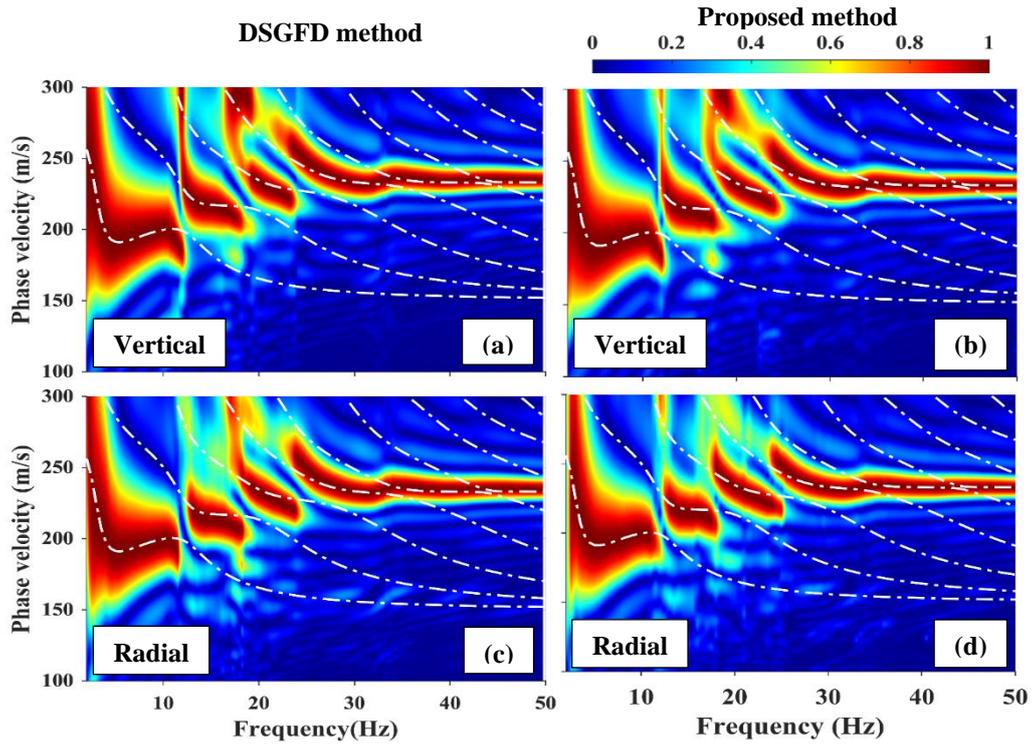

**Fig. 10.** Comparison of dispersion images obtained using proposed method and DSGFD method for Profile III: (a) vertical component from DSGFD method, (b) vertical component from proposed method, (c) radial component from DSGFD method, and (d) radial component from the proposed method.

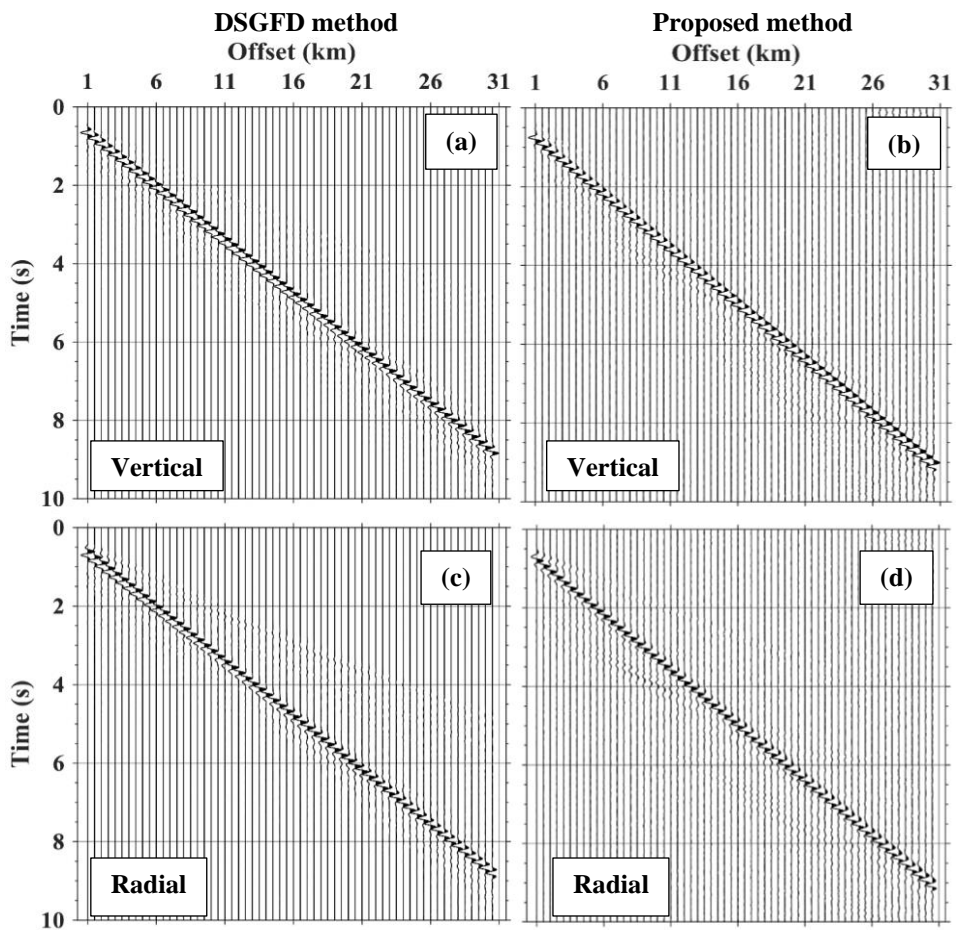

**Fig. 11.** Comparison of seismograms obtained using proposed method and DSGFD method for Profile IV: (a) vertical component from DSGFD method, (b) vertical component from proposed method, (c) radial component from DSGFD method, and (d) radial component from the proposed method.

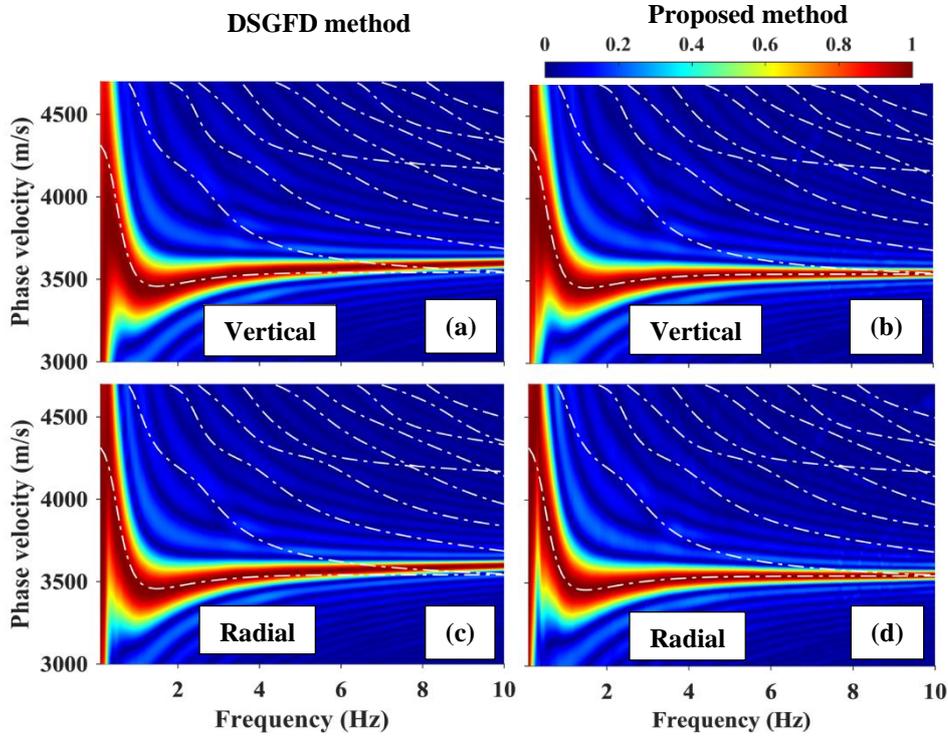

**Fig. 12.** Comparison of dispersion images obtained using proposed method and DSGFD method for Profile IV: (a) vertical component from DSGFD method, (b) vertical component from proposed method, (c) radial component from DSGFD method, and (d) radial component from the proposed method.

### 3.2 Computational Advantages

In the previous section, the accuracy of the proposed approach was demonstrated across different soil models. Since the method assumes lateral homogeneity of the medium, it only requires vertical discretization. Furthermore, implementing higher-order interpolation functions enhances the proposed approach's accuracy and computational efficiency. Fig. 13 provides a comparison of computational times for the four representative models: the traditional SGFD method, the DSGFD method, the HTLM wavefield modeling approach proposed in Bhaumik and Naskar (2023a), and the present method. All programs were run in MATLAB R2023a on a computer with 16 GB RAM and a clock speed of 3.2 GHz. The DSGFD MATLAB programs were coded according to the method described in Bhaumik and Naskar [27]. In all the cases, the computational time of the present method is similar to the HTLM-based wavefield modeling approach. For profile I, the proposed technique is more than 440 times faster than the SGFD method and 340 times faster than the DSGFD method. The traditional SGFD approach in Profile II is computationally intensive, taking over 120 minutes. This extended computational time for Profile II is primarily attributed to the necessary spatial and temporal resolutions. A 0.1 m spatial grid resolution is necessary for precisely modeling layer interfaces in Profile II. At the same time, the small temporal resolution is mandated by the high compression wave velocity of the medium. In contrast, the DSGFD method requires 85 minutes, a substantial reduction compared to the SGFD approach. However, the proposed method demonstrates a remarkable efficiency, requiring only 0.8 seconds. For the modeling of Profile III, the proposed method is 240 times faster than the SGFD method and 190 times faster than the DSGFD method. Finally, in the case of Profile IV, the proposed method surpasses both traditional approaches by approximately 1000 times in terms of computational speed. These results underscore the significant computational advantages the proposed method offers, making it a highly efficient choice for modeling a variety of geological profiles. It's worth noting that Poisson's ratio influences the computational time of time-domain numerical methods. Materials with high Poisson's ratios have higher compression wave velocities, necessitating smaller time steps and resulting in longer computational times. Bhaumik and Naskar [16] have discussed the relationship between computational time and Poisson's ratio for numerical methods. Given that the proposed method operates in the frequency domain, its computational time is nearly insensitive to compression wave velocity. Consequently, the proposed method is well-suited for near-surface models featuring saturated layers or a water table close to the surface, where computational efficiency is paramount.

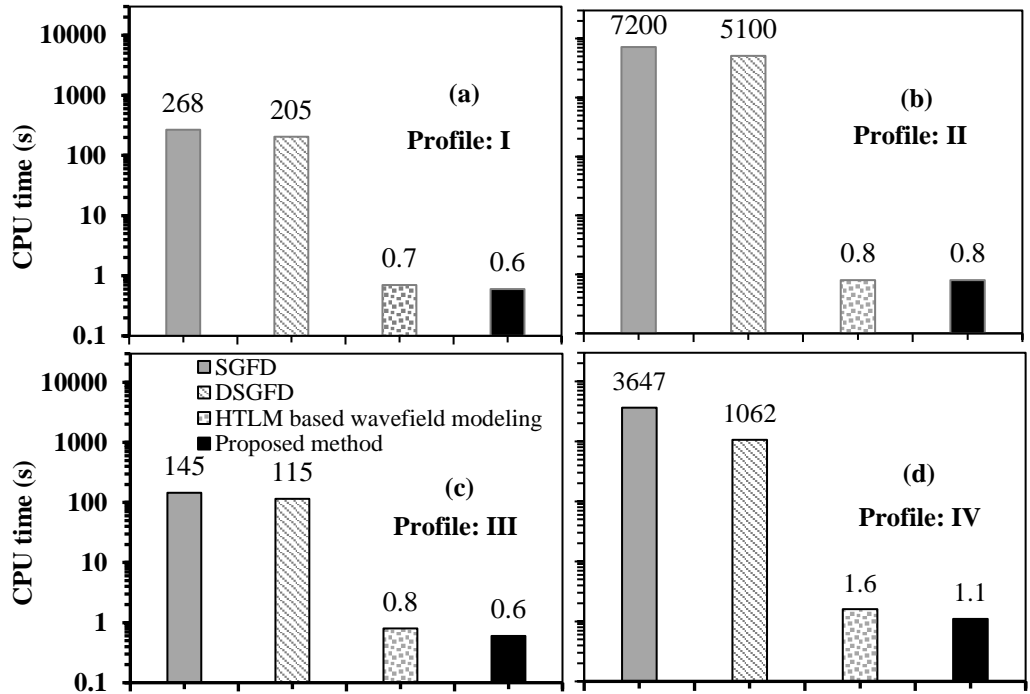

**Fig. 13**. Comparison of computational time among the proposed method, SGFD method, DSGFD method and the HTLM based wavefield modelling approach proposed in Bhaumik and Naskar (2023a).

### 3.3 Influence of Leaky Waves

While modeling soil profiles characterized by a high Poisson's ratio, it produces leaky waves [32]. These waves travel at velocities higher than the maximum shear wave velocity of the medium yet lower than the compression wave velocity. Theoretical solutions of the governing equations for a homogeneous half-space yield three roots. When the Poisson's ratio is less than 0.26, only one of these roots satisfies the criteria for a surface wave. However, when the Poisson's ratio exceeds 0.26, the solution produces one real root and two complex conjugate roots [33]. The real roots correspond to the phase velocity of the Rayleigh wave, while the complex roots represent the leaky waves, which propagate at velocities higher than the medium's shear wave velocity [32,33]. It is important to note that a higher Poisson's ratio results in greater contamination by these leaky waves. Traditional modeling approaches, such as simple mode summation, cannot accurately represent these leaky waves. The proposed approach considers both the real and complex roots, enabling it to effectively model the behavior of these leaky waves in soil profiles characterized by high Poisson's ratios. While processing techniques are available, such as muting and removal [32], to mitigate the risk of misidentifying leaky modes as part of the Rayleigh wave mode, eliminating them entirely from the wavefield is challenging. Using numerical techniques, the influence of leaky waves cannot be avoided. However, in the proposed approach, it is possible to mitigate this influence by limiting the eigenvalues to those corresponding to the maximum shear wave velocity of the medium and ignoring the complex roots associated with leaky waves. Fig 14 shows the dispersion curves obtained for different Poisson's ratios in Profile I. For Poisson's ratio 0.3, the modeled dispersion curve closely aligns with the theoretical one, reflecting a lesser influence of leaky waves at this value (Fig. 14a). Fig. 14b illustrates the peak dispersion curves obtained by limiting the eigenvalues. However, increasing the Poisson's ratio to 0.4, clear evidence of overestimating phase velocity is observed below 7 Hz (Fig. 14c). This deviation becomes more pronounced with further increases in the Poisson's ratio to 0.49 (Fig. 14e). It is worth highlighting that, across all these cases, the dispersion curve modeled using the proposed approach consistently matches well with the one obtained using the DSGFD method. Meanwhile, Fig. 14d and f illustrate the peak dispersion curves obtained by limiting the eigenvalues. The modeled dispersion curve closely aligns with the theoretical one, neglecting the leaky wave effect, even for a high Poisson's ratio. This demonstrates that the proposed forward modeling approach can be effectively employed across various scenarios, providing a valuable tool for accurately characterizing dispersion curves.

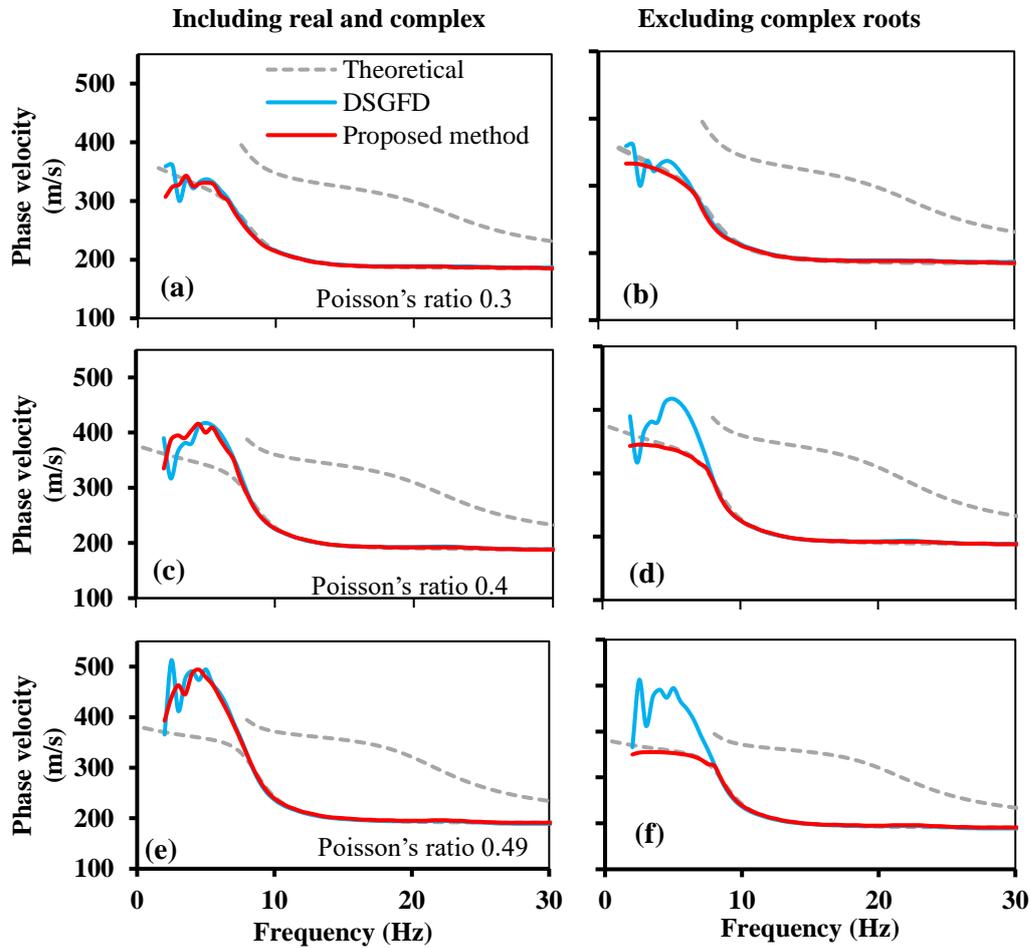

**Fig. 14.** Dispersion curve for Profile I with different Poisson's ratios, considering all the real and complex eigenvalues to model leaky waves (left column), and with excluding the complex roots to eliminate the leaky wave effect (right column). Panels (a)-(b) represent a Poisson's ratio of 0.3, (c)-(d) a Poisson's ratio of 0.4, and (e)-(f) a Poisson's ratio of 0.49.

### 3.4 Influence of Source Offset

In active surface wave-based geotechnical field tests, placing the source close to receivers results in a cylindrically propagating wavefront and introduces body waves in the measured wavefield[34]. The mode activated by the source exhibits velocities that deviate from the characteristics of planar waves, particularly in the vicinity of the source location[34]. The area where the assumption of planer wavefront is not valid is referred to as the near-field region. The underestimation and oscillation of Rayleigh wave phase velocity near the low frequencies is the primary consequence of the near-field effect [34,35]. These discrepancies observed in the dispersion curve propagate errors in the inversion analysis. Therefore, it is crucial to investigate the capability of the forward model to capture the source offset effects. The inclusion of both propagating modes with real wavenumbers and decaying modes with complex wavenumbers enables accurate computation of surface displacement not only in the far-field but also close to the source[19,24]. Furthermore, including cylindrical waveforms in the form of Hankel functions better captures the field scenario. The method's effectiveness in capturing discrepancies at low frequencies is evaluated by studying Profile-I's source offset and spread length effects. A Poisson's ratio of 0.26 is maintained for both the layer and half-space to distinguish from the contamination of leaky waves. The dispersion spectrum obtained from the proposed method is compared to those derived from planer wave-based HTLM wavefield modeling [16] and the DSGFD method. Fig. 15 illustrates the dispersion spectrum corresponding to 24 m spread lengths, with source offsets of 1 m, 5 m, 10 m, and 20 m. With 1 m source offset, the modeled dispersion spectrum using the planar wave-based HTLM approach aligns well with the theoretical fundamental mode over the entire frequency range. However, the spectral energy obtained by the DSGFD method and the proposed method deviates from the theoretical curve and underestimates the phase velocity below 8 Hz (Fig.15 b, c). The near-field effects result from the shorter source offset and limited spread length, leading to underestimated phase velocities in both the present approach and the DSGFD method. This effect can be alleviated by positioning the receiver aperture at a greater distance from the source. Fig. 15d, e, f illustrates the dispersion spectrum with source-to-1st-sensor distances of 5 m. It can be observed that the phase velocity below 6 Hz is affected by the near-field effect in both the present method and the DSGFD method. However, the underestimation of phase velocity is less compared to the 1 m offset. Increasing the near offset distance to 10 m reduces the impact of the near-field effect (Fig. 15g, h, i), and with a 20 m offset, the spectrum closely matches the planar wave-based dispersion image (Fig. 15j, k, l).

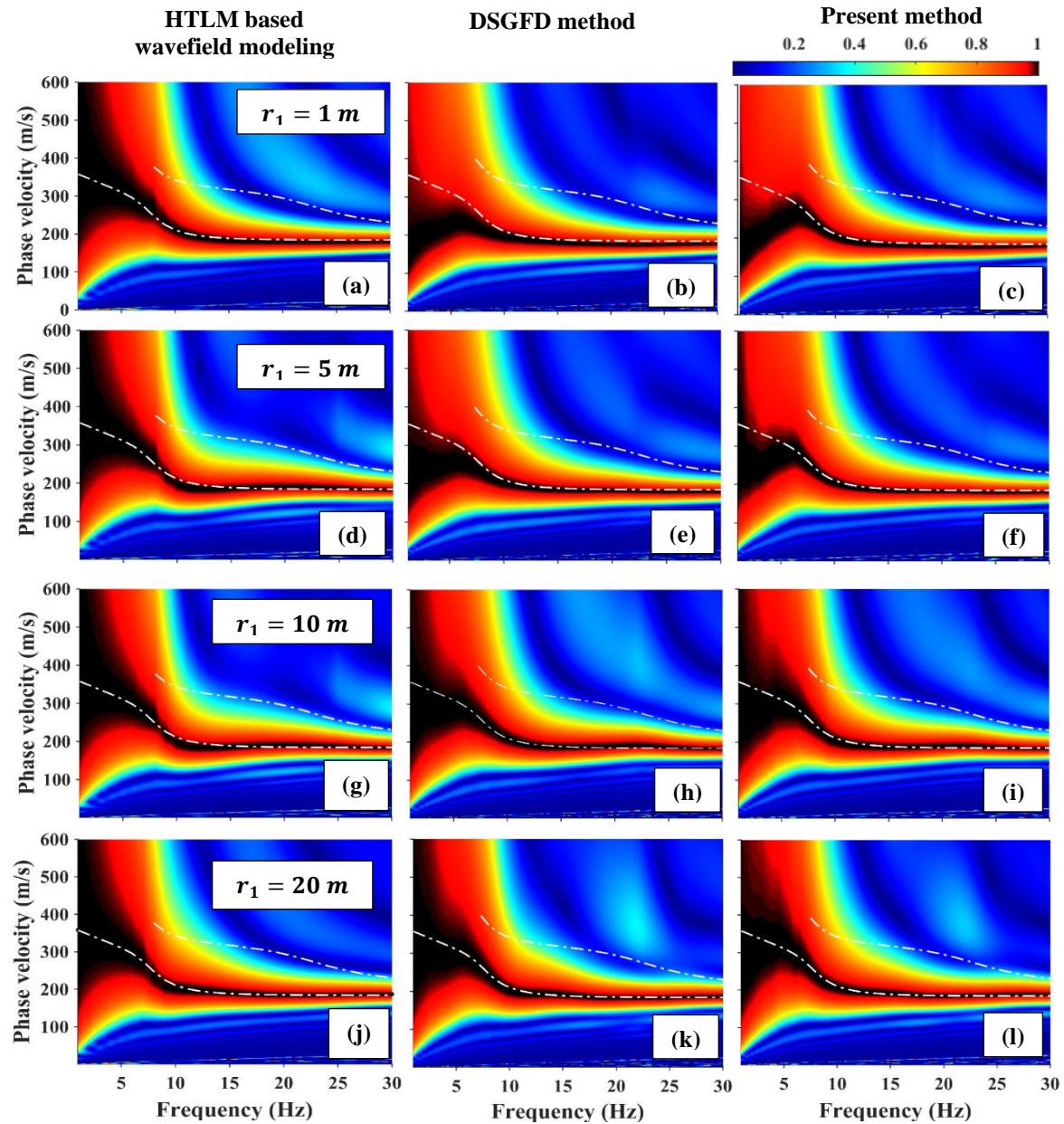

**Fig. 15.** Comparison of dispersion spectrum for Profile I with different source offsets using the DSGFD method, HTLM-based wavefield modeling and proposed method ($X = 24$, $dx = 1\ m$). Panels (a)-(c) correspond to 1 m offset, (d)-(f) 5 m offset; (g)-(i) 10 m offset, and (j)-(l) 20 m offset. Thin white dash-dot lines represent the first two modes.

Overall, the dispersion images generated by the proposed approach for various offsets closely match the dispersion phenomena observed in the DSGFD method. The minor discrepancy in the low frequency between the numerical and present method can be attributed to the fundamental difference between the two approaches: the 2D discretized DSGFD operates in the time domain, whereas the proposed method works in the frequency domain. Additionally, in the calculation of the dispersion spectrum using limited receivers, the frequency domain surface responses are inevitably truncated, leading to distortions and overestimation in the spectral values at low frequencies [36]. However, this discrepancy was observed to diminish with an increase in the receiver spread length and source offset distance.

### 3.5 Modeling Lamb Waves

Lamb waves have practical applications in various engineering fields, including ultrasonic testing, non-destructive evaluation (NDE), and structural health monitoring. Lamb waves exhibit a dispersive nature and are highly sensitive to factors like defects or changes in material properties such as thickness, stiffness, or density. Investigating Lamb waves within layered structures offers valuable insights into wave interactions, encompassing mode conversion and energy transmission mechanisms. The proposed method allows for the computation of the dispersion spectrum of Lamb wave modes in plate-like structures. There is no half-space, so PMDL elements are not required to model plate-like structures. To demonstrate the effectiveness of the proposed approach in modelling Lamb waves, a 200 mm concrete plate with a shear wave velocity of 2485 m/s, Poission's ratio of 0.2, and density of 2400 kg/m$^3$ is adopted from Lin et al. (2022b). The dispersion spectrum was generated using 48 sensors spaced at 0.05 m. Fig. 16 presents a comparison of the dispersion curves obtained using the proposed approach, numerical solutions [37], and experimental data [38] in the frequency range of 0 to 30 kHz. Remarkably, the dispersion trends obtained using the proposed approach closely match the numerical and experimental results. The dominance of the A0 mode is observed in both the experimental and modeled curves up to 9000 Hz, followed by an energy jump to the S0 mode. While the experimental results indicate a shift in dominance to the A0 mode from 10000 Hz, the energy trend obtained through the current method aligns reasonably well with the numerical results from Lin et al. (2022b). It is important to acknowledge that the slight mismatch is attributed to uncertainties in the material parameters considered for the analysis.

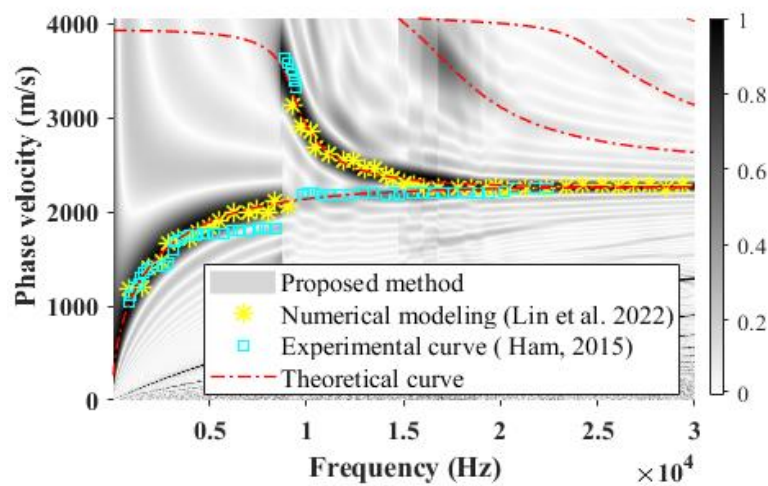

**Fig. 16.** Comparison of dispersion spectrum of Lamb waves propagating through a 200 mm concrete slab.

### 4 CONCLUSION

This study presented an active source-based semi-analytical wavefield modeling approach to develop a dispersion spectrum of Rayleigh waves propagating through layered half-space. The proposed framework models the wave propagation considering cylindrical wavefront in the form of Hankel function instead of planer wave assumption. It incorporates propagating waves with real wavenumbers and decaying waves with complex wavenumbers, thereby modeling surface responses in both the near-field and far-field. The method calculates both the vertical and radial component response of the Rayleigh wave generated from a vertical source at the surface. Higher order thin layer method was employed to calculate element stiffness matrices, and perfectly matched discrete layers were used to model the elastic half-space. The efficacy of the proposed formulation is demonstrated using four different soil profiles, including regularly dispersive, irregular dispersive, and crustal-scale models. The Rayleigh wave vertical and radial component dispersion images for all representative soil profiles match well with the reference solution obtained by staggered grid finite difference-based 2D wavefield modeling. The results of the different soil models illustrated that the proposed framework effectively models complex phenomena such as mode jumping and mode osculation. The radial component dispersion image is especially beneficial in the presence of significant impedance contrast or a rock bed. Furthermore, the proposed method effectively captures the smooth transition of modal energy from the fundamental mode to higher modes at low frequencies, addressing the challenges of mode misidentification and overestimation of shear wave velocity during inversion. The method produces a dispersion spectrum which is sensitive to the active source location. It demonstrates the underestimation of phase velocity at low frequency while placing the receivers close to the source due to the near-field effect. The proposed forward modeling approach can effectively model leaky waves in media with a high Poisson's ratio. Moreover, it has the capability to control the contamination of leaky waves by limiting the complex wavenumbers. The application of the proposed formulation was extended to model Lamb waves propagating through plate structures. The computational time of the presented method is at least two orders of magnitude faster than the reference numerical analysis. Depending on the model type, where the 2D numerical method takes several minutes, the present approach generates results within seconds. Furthermore, unlike 2D numerical methods, the proposed method's computational time is insensitive to Poisson's ratio. As the proposed technique preserves all the properties of a complete wavefield, including the near-field effect and higher modes, it can effectively be used to develop an advanced inversion algorithm where the entire spectrum is considered. The present approach will help to generate a large training dataset for convolutional neural network (CNN) based inversion analysis.

**APPENDIX A**

The global matrices $\mathbf{A}, \mathbf{B}, \mathbf{C}$, and $\mathbf{M}$ are obtained by assembling all the layer stiffness matrices, $\mathbf{A}_j, \mathbf{B}_j, \mathbf{C}_j$ and $\mathbf{M}_j$. Here, the displacement vectors are arranged by first grouping all the horizontal displacements, followed by the vertical displacements. Accordingly, the layer matrices for the layer $j$ can be expressed in the form:

$$\mathbf{A}_j = \begin{bmatrix} \mathbf{A}_{xx_j} & 0 \\ 0 & \mathbf{A}_{zz_j} \end{bmatrix}, \quad \mathbf{B}_j = \begin{bmatrix} 0 & \mathbf{B}_{xz_j} \\ -\mathbf{B}_{zx_j} & 0 \end{bmatrix},$$

$$\mathbf{G}_j = (\mathbf{C}_j - \omega^2 \mathbf{M}_j) = \begin{bmatrix} \mathbf{C}_{xx_j} - \omega^2 \mathbf{M}_{xx_j} & 0 \\ 0 & \mathbf{C}_{zz_j} - \omega^2 \mathbf{M}_{zz_j} \end{bmatrix} = \begin{bmatrix} \mathbf{G}_{xx_j} & 0 \\ 0 & \mathbf{G}_{zz_j} \end{bmatrix} \quad \text{(A-1)}$$

in which,

$$\mathbf{A}_{xx_j} = (\lambda_j + 2\mu_j) \int_{z_j}^{z_{j+1}} \mathbf{N}^T \mathbf{N} dz, \qquad \mathbf{A}_{zz_j} = \mu_j \int_{z_j}^{z_{j+1}} \mathbf{N}^T \mathbf{N} dz,$$

$$\mathbf{B}_{xz_j} = \lambda_j \int_{z_j}^{z_{j+1}} \mathbf{N}^T \mathbf{N}' dz - \mu_j \int_{z_j}^{z_{j+1}} \mathbf{N}'^T \mathbf{N} dz, \qquad \mathbf{B}_{zx_j} = \mathbf{B}_{xz_j}^T,$$

$$\mathbf{C}_{xx_j} = \mu_j \int_{z_j}^{z_{j+1}} \mathbf{N}'^T \mathbf{N}' dz, \qquad \mathbf{C}_{zz_j} = (\lambda_j + 2\mu_j) \int_{z_j}^{z_{j+1}} \mathbf{N}'^T \mathbf{N}' dz, \quad \text{(A-2)}$$

$$\mathbf{M}_{xx_j} = \rho_j \int_{z_j}^{z_{j+1}} \mathbf{N}^T \mathbf{N} dz, \qquad \mathbf{M}_{zz_j} = \rho_j \int_{z_j}^{z_{j+1}} \mathbf{N}^T \mathbf{N} dz$$

in which $\lambda_j$ and $\mu_j$ are Lame's parameters. The shape function Kronecker product $\mathbf{N} = \overline{\mathbf{N}} \otimes \mathbf{I}_{2\times 2}$ where, $\overline{\mathbf{N}} = [\overline{N_1}, \dots, \overline{N_e}]$ is shape function vector for 1D $e$ noded element. The original TLM matrix elements introduced in Kausel and Roësset [2] are derived based on two noded linear finite elements. This approach requires a very fine discretization of each layer to achieve the desired level of accuracy. However, higher-order finite elements can provide better accuracy than linear elements. In this study, higher-order shape functions or interpolation functions are used to construct the layer stiffness matrices, $\mathbf{A}_j, \mathbf{B}_j, \mathbf{C}_j$ and $\mathbf{M}_j$. The higher order 1D shape function is derived from Lagrange interpolation:

$$\bar{N}_b(z) = \prod_{\substack{a=1 \\ a \neq b}}^{e} \frac{z - z_a}{z_b - z_a} \tag{A-3}$$

where, $a$ and $b$ are the data points. To calculate the contribution matrices in equation A-2, the numerical integration can be easily performed by using Gauss quadrature with an appropriate pair of nodes and their corresponding weights.

**FIGURE CAPTIONS**

**Fig. 1.** (a) Layered half-space model (b) discretization of finite layers and infinite half-space.

**Fig. 2**. Illustration of cylindrical and planer waveform.

**Fig. 3.** Displacement response due to circular plate loading.

**Fig. 4**. Four representative synthetic soil profiles.

**Fig. 5.** Comparison of seismograms obtained using proposed method and DSGFD method for Profile I: (a) vertical component from DSGFD method, (b) vertical component from proposed method, (c) radial component from DSGFD method, and (d) radial component from the proposed method.

**Fig. 6**. Comparison of dispersion images obtained using proposed method and DSGFD method for Profile I: (a) vertical component from DSGFD method, (b) vertical component from proposed method, (c) radial component from DSGFD method, and (d) radial component from the proposed method.

**Fig. 7.** Comparison of seismograms obtained using proposed method and DSGFD method for Profile II: (a) vertical component from DSGFD method, (b) vertical component from proposed method, (c) radial component from DSGFD method, and (d) radial component from the proposed method.

**Fig. 8.** Comparison of dispersion images obtained using proposed method and DSGFD method for Profile II: (a) vertical component from DSGFD method, (b) vertical component from proposed method, (c) radial component from DSGFD method, and (d) radial component from the proposed method.

**Fig. 9.** Comparison of seismograms obtained using proposed method and DSGFD method for Profile III: (a) vertical component from DSGFD method, (b) vertical component from proposed method, (c) radial component from DSGFD method, and (d) radial component from the proposed method.

**Fig. 10.** Comparison of dispersion images obtained using proposed method and DSGFD method for Profile III: (a) vertical component from DSGFD method, (b) vertical component from proposed method, (c) radial component from DSGFD method, and (d) radial component from the proposed method.

**Fig. 11.** Comparison of seismograms obtained using proposed method and DSGFD method for Profile IV: (a) vertical component from DSGFD method, (b) vertical component from proposed method, (c) radial component from DSGFD method, and (d) radial component from the proposed method.

**Fig. 12.** Comparison of dispersion images obtained using proposed method and SGFD method for Profile IV: (a) vertical component from SGFD method, (b) vertical component from proposed method, (c) radial component from SGFD method, and (d) radial component from the proposed method.

**Fig. 13**. Comparison of computational time among the proposed method, SGFD method, DSGFD method and the HTLM based wavefield modelling approach proposed in Bhaumik and Naskar [15].

**Fig. 14.** Dispersion curve for Profile I with different Poisson's ratios, considering all the real and complex eigenvalues to model leaky waves (left column), and with excluding the complex roots to eliminate the leaky wave effect (right column). Panels (a)-(b) represent a Poisson's ratio of 0.3, (c)-(d) a Poisson's ratio of 0.4, and (e)-(f) a Poisson's ratio of 0.49.

**Fig. 15.** Comparison of dispersion spectrum for Profile I with different source offsets using the DSGFD method, HTLM-based wavefield modeling and proposed method ($X = 24$, $dx = 1\ m$). Panels (a)-(c) correspond to 1 m offset, (d)-(f) 5 m offset; (g)-(i) 10 m offset, and (j)-(l) 20 m offset. Thin white dash-dot lines represent the first two modes.

**Fig. 16.** Comparison of dispersion spectrum of Lamb waves propagating through a 200 mm concrete slab.

**TABLE CAPTIONS**

**Table 1.** Layer parameters of adopted soil models